\documentclass[prc,floatfix,groupedaddress,nofootinbib,showpacs,preprintnumbers,
amsmath,amssymb,amsfonts,superscriptaddress,widetable] {revtex4-1}
\usepackage{graphicx}
\usepackage{dcolumn}
\usepackage{mathrsfs,amsmath,amssymb}
\usepackage{bm}
\usepackage{graphicx}
\usepackage{dcolumn}
\usepackage{bm}
\usepackage{array}
\usepackage{bigstrut}
\usepackage[usenames]{color}
\begin{document}

\title{Nuclear Mass Predictions for the Crustal Composition of Neutron Stars: \\
A Bayesian Neural Network Approach}
\author{R. Utama}
\email{ru11@my.fsu.edu} 
\affiliation{Department of Physics, Florida State University, Tallahassee, FL 32306} 
\author{J. Piekarewicz}
\email{jpiekarewicz@fsu.edu}
\affiliation{Department of Physics, Florida State University, Tallahassee, FL 32306}
\author{H. B. Prosper}
\email{harry@hep.fsu.edu}
\affiliation{Department of Physics, Florida State University, Tallahassee, FL 32306}
\date{\today}
\begin{abstract}
\begin{description}
\item[Background] Besides their intrinsic nuclear-structure value, nuclear mass 
         models are essential for astrophysical applications, such as r-process
         nucleosynthesis and neutron-star structure. 
\item[Purpose] To overcome the intrinsic limitations of existing ``state-of-the-art'' 
	 mass models through a refinement based on a Bayesian Neural Network 
	 (BNN) formalism.  
\item[Methods] A novel BNN approach is implemented with the goal of optimizing 
        mass residuals between theory and experiment.
\item[Results] A significant improvement (of about 40\%) in the mass predictions of
	 existing models is obtained after BNN refinement. Moreover, these improved 
	 results are now accompanied by proper statistical errors. Finally, by constructing 
        a ``world average'' of these predictions, a mass model is obtained that is used 
        to predict the composition of the outer crust of a neutron star. 
\item[Conclusions] The power of the Bayesian neural network method has been
        successfully demonstrated by a systematic improvement in the accuracy of
        the predictions of nuclear masses. Extension to other nuclear observables 
        is a natural next step that is currently under investigation.
\end{description}
\end{abstract}
\maketitle

\section{Introduction}
\label{intro}

Shortly after the discovery of the neutron by Chadwick, the remarkable
semi-empirical nuclear mass formula of Bethe and Weizs\"acker was 
conceived. Originally proposed by Gamow and later extended by 
Weizs\"acker, Bethe, Bacher, and others\,\cite{Weizsacker:1935,Bethe:1936}, 
the  ``liquid-drop'' model (LDM) regards the nucleus as an incompressible drop 
consisting of two quantum fluids, one electrically charged consisting of $Z$ 
protons and one neutral containing $N$ neutrons. Given that the nuclear 
binding energy $B(Z,N)$ accounts for only a small fraction ($\lesssim$\,1\%) 
of the total mass of the nucleus, it is customary to remove the large, but 
well known, contribution from the mass of its constituents. That is,
\begin{equation}
 B(Z,N)\!\equiv\!Zm_{p}\!+\!Nm_{n}\!-\!M(Z,N),
 \label{BE}
\end{equation}
where $A\!=\!Z\!+\!N$ is the mass (or baryon) number of the nucleus.
In this manner $B(Z,N)$ encapsulates all the complicated nuclear
dynamics. In the context of the liquid-drop formula, the binding energy 
is written in terms of a handful of empirical parameters that represent
volume, surface, Coulomb, asymmetry, and pairing contributions:
\begin{equation}
 B(Z, A) = a_{\rm v} A - a_{\rm s} A^{2/3} - a_{\rm c} \frac{Z^2}{A^{1/3}} - 
 \Big(a_{\rm a}+\frac{a_{\rm as}}{{A^{1/3}}}\Big)\frac{(A-2Z)^2}{A}- 
 a_{\rm p}\frac{\eta{(Z,N)}}{A^{1/2}} + \ldots
\label{LDM}  
\end{equation} 
where the pairing coefficient takes values of $\eta\!=$+1,0,-1 depending
on whether an even-even, even-odd, or odd-odd nucleus is involved. Note
that besides the conventional volume asymmetry term, a surface asymmetry term has 
also been included\,\cite{Nikolov:2010dp}. The handful of empirical coefficients 
are determined through a least-squares fit to the thousands of nuclei whose 
masses have been determined accurately\,\cite{AME:2012}. It is indeed a 
remarkable fact that in spite of its enormous simplicity the 80 year old LDM 
has stood the test of  time. 

To a large extent, the reason that the LDM continues to be enormously 
valuable even today is because the dominant contribution to the nuclear 
binding energy varies smoothly with both $Z$ and $N$. Indeed, according 
to Strutinsky's energy theorem\,\cite{Strutinsky:1967}, the nuclear binding 
energy may be separated into two main components: one large and smooth 
and another one small and fluctuating. Whereas successful in reproducing 
the smooth general trends, the LDM fails to account for the rapid fluctuations 
with $Z$ and $N$ around shell gaps. The explanation for the extra stability 
observed around certain ``magic numbers'' had to await the 
insights of Haxel, Jensen, Suess, and 
Goeppert-Mayer\,\cite{Haxel:1949,Mayer:1950b}, who elucidated the vital 
role of the spin-orbit interaction in nuclear physics. Since the seminal work
by Goeppert-Mayer and Jensen, who shared with Wigner the 1963 Nobel 
Prize, theoretical calculations have evolved primarily along two separate 
lines of investigation. One of them---the so-called microscopic-macroscopic 
(``mic-mac'') model---incorporates microscopic corrections 
to account for the physics that is missing from the most 
sophisticated macroscopic models. Mic-mac approaches have enjoyed 
their greatest success in the work of M\"oller 
\emph{et al.}\,\cite{Moller:1981zz,Moller:1988,Moller:1993ed} and 
Duflo and Zuker\,\cite{Duflo:1995}. 
The second theoretical approach, falling under the general classification 
of microscopic mean-field models, relies on an energy density functional 
that is motivated by well known features of the nuclear dynamics. Such
density functionals are expressed in terms of a handful of empirical 
constants that are directly fitted to experimental 
data\,\cite{Goriely:2010bm,Kortelainen:2010hv,Erler:2012qd,Chen:2014sca}. 

Theoretical models of nuclear masses such as the ones discussed 
above are of critical importance in our quest to understand the nature 
of the strong nuclear force. Fundamental questions at the core of 
nuclear structure include: How do magic numbers evolve as one 
moves away from stability? What are the limits of nuclear existence? 
How does one access the purported island of stability of superheavy nuclei? 
Besides their prominent place in nuclear structure, nuclear masses also 
play a vital role in understanding a variety of astrophysical phenomena, 
such as r-process nucleosynthesis and the composition of the neutron-star 
crust. Unfortunately, answers to all of these critical questions are hindered 
by the need to extrapolate to uncharted regions of the nuclear landscape. 
Indeed, whereas model predictions tend to agree near stability, they are 
often in stark disagreement far away from their region of applicability;
see, for example, Fig.\,42 in Ref.\,\cite{Blaum:2006}.

Given the critical importance of nuclear masses in elucidating certain 
astrophysical phenomena, the search for an alternative approach to 
compute nuclear masses is justified, perhaps even at the expense of 
sacrificing some physical insights. Falling within this category are the 
Garvey-Kelson relations (GKRs), which are based on two \emph{local} 
mass relations each involving six neighboring 
nuclei\,\cite{Garvey:1966zz,GARVEY:1969zz}. As such, the GKRs 
may be used to predict the mass of an unknown nuclide in terms of its 
known neighbors. The half-century old GKRs have been recently revitalized 
because of an interest in understanding any inherent limitation in 
nuclear-mass models\,\cite{Barea:2005fz,Barea:2008zz,Morales:2009pq}. 
Shortly thereafter, and guided by Strutinsky's energy theorem, valuable insights 
into the underlying success of the GKRs were 
developed\,\cite{Piekarewicz:2009av}. In particular, it was shown that the 
validity of the GKRs requires that derivatives of the underlying mass 
function $M(Z,N)$ of third order and higher  
vanish\,\cite{Preston:1993,Piekarewicz:2009av}. Given that 
successive derivatives of any smooth function are progressively smaller, 
the GKRs are well satisfied by the large and smooth contribution of the 
underlying mass function. Moreover, the GKRs were constructed in such 
a way that all residual two-body interactions that enter into mass relations 
are exactly cancelled to first 
order\,\cite{Garvey:1966zz,GARVEY:1969zz,Preston:1993}. Although not 
rooted in firm fundamental physical principles, the GKRs predictions rival 
some of the most successful mass formulae 
available in the literature\,\cite{Barea:2008zz,Morales:2009pq}. Finally, given 
that the success of the GKRs hinges on an underlying smooth mass function, 
it was concluded that the formalism could be suitably extended to other 
physical observables that display similar behavior, such as nuclear charge 
radii\,\cite{Piekarewicz:2009av}.  

In this contribution we continue to rely on Strutinsky's energy theorem for
the implementation of a novel Bayesian Neural Network (BNN) approach to 
the calculation of nuclear masses; see Ref.\,\cite{Athanassopoulos:2003qe}
for the use of neural networks in the study of nuclear mass systematics and
Ref.\,\cite{Neal1996} for a general exposition.
However, unlike the Garvey-Kelson 
relations, the present approach offers a global description of nuclear 
masses. To introduce the method we adopt a simple liquid-drop formula 
to describe the large and smooth contribution of the underlying mass
function $M_{\rm LDM}(Z,N)$. To account for the small and fluctuating 
contribution, we ``train'' a suitable neural network on the 
\emph{mass residuals} between the LDM predictions and experiment,
as given in the latest Atomic Mass Evaluation (AME2012)\,\cite{AME:2012}. 
Once trained, we used the resulting ``universal approximator" 
$\delta_{\rm LDM}(Z,N)$ to validate the approach and later 
to make predictions in regions where experimental data are
unavailable. That is, the resulting mass formula becomes
\begin{equation}
 M(Z,N) \equiv M_{\rm LDM}(Z,N) + \delta_{\rm LDM}(Z,N).
 \label{BNN_LDM}
\end{equation}
The underlying philosophy behind our implementation of the BNN approach 
is to incorporate as much physics as possible in the choice of the large and 
smooth component and then relinquish control to a sophisticated numerical 
algorithm to model the small and fluctuating part. However, note that 
although inspired by such a concept, the proposed approach goes beyond 
Strutinsky's energy theorem. For example, the main component of the mass 
formula may already include---at least in part---the small and fluctuating
component (for example, by using the mass formula of Duflo and Zuker).
Thus, the BNN approach is left with the task of performing the fine tuning.
Finally, given that the predictions of the residuals involve the calibration of 
a universal approximator constructed using a Bayesian method, 
all mass predictions are accompanied by properly 
estimated theoretical errors.

As a concrete application of the BNN method, we explore the role of nuclear 
masses on the composition of the outer crust of a neutron star. At the 
densities of relevance to the outer crust, the average inter-nucleon separation 
is considerably larger than the range of the nuclear interaction. Thus, it is 
energetically favorable for nucleons to cluster into individual nuclei that, in
turn, arrange themselves in a crystalline lattice. This crystalline lattice is
itself immersed in a uniform free Fermi gas of electrons that are critical to 
maintain the overall charge neutrality of the crust\,\cite{Baym:1971pw}. 
Although the dynamics of the outer crust is relatively simple, its composition 
is highly sensitive to the nuclear mass model\,\cite{RocaMaza:2008ja}. For 
example, at the top layers of the crust where the density is extremely low 
($\sim\!10^{4}\,{\rm g/cm^{3}}$) the crystal lattice is composed of 
${}^{56}$Fe nuclei---the nucleus with the lowest mass per nucleon in the 
nuclear chart. However, as the density increases, ${}^{56}$Fe ceases to 
be the preferred nucleus. This is because the electronic contribution to 
the total energy increases rapidly with density. Thus, in an effort to 
minimize the overall energy of the system, it becomes advantageous for 
the electrons to capture on protons, thereby making the preferred nucleus 
more neutron rich. As the density continues to increase, the crustal composition 
evolves into a Coulomb lattice of progressively more exotic neutron-rich nuclei. 
Finally, at a density of about $4\!\times\!10^{11}\,{\rm g/cm^{3}}$ (still about 
three orders of magnitude below nuclear matter saturation density) the neutron 
drip line is reached. Although most mass models predict that this sequence 
of progressively more exotic nuclei terminates with ${}^{118}$Kr 
($Z\!=\!36$ and $N\!=\!82$), it is worth noting that the last isotope with a 
well measured mass is ${}^{97}$Kr---21 neutrons away from ${}^{118}$Kr. 
Hence, the reliance on mass models that are often hindered 
by uncontrolled extrapolations is, unfortunately, unavoidable. However, we
are at the dawn of a new era where rare isotope facilities will probe the limits 
of nuclear existence and in so doing will provide critical guidance to theoretical
models. Indeed, a recent landmark experiment at ISOLTRAP/CERN 
measured for the first time the mass of the ${}^{82}$Zn isotope\,\cite{Wolf:2013ge}. 
Owing to the sensitivity of the crustal composition to the mass model, it was found 
that the addition of this one mass value alone resulted in an interesting 
modification to the composition of the outer crust\,\cite{Wolf:2013ge,Pearson:2011zz}. 

It is the aim of this contribution to use a BNN approach to create a global
mass model that may be used to examine the composition of the outer crust. 
This challenging task involves knowledge of nuclear masses along three 
separate regions of the nuclear chart. The first region impacts the top layers of 
the outer crust where the density is at its lowest. In this region the electronic 
contribution to the energy is moderate, so the isotopes of relevance are located 
around the stable iron-nickel region where the nuclear masses are very accurately 
known. The second region of interest involves nuclei around the $N\!=\!50$ magic 
number; typically from Zr ($Z\!=\!40$) to Ni ($Z\!=\!28$). This region lies at the 
border between accurately known masses (such as in the case of ${}^{90}$Zr, 
${}^{88}$Sr, and ${}^{86}$Kr) and poorly constrained masses of very 
neutron-rich nuclei (such as ${}^{78}$Ni and until very recently ${}^{82}$Zn).  
Given that there is some experimental information available in this 
region, local methods such as the Garvey-Kelson relations may provide 
reliable estimates for the masses that have yet to be measured. The third
and last region involves nuclei around neutron magic number $N\!=\!82$
where little or no experimental information is available. Depending on the
mass model, the nuclei of relevance span the region from ${}^{132}$Sn 
($Z\!=\!50$) all the way down to ${}^{118}$Kr\,\cite{RocaMaza:2011pk}. 
Clearly, local methods such as the Garvey-Kelson relations are of very
limited use. Thus, in this contribution we attempt to construct a global 
mass model by relying on a BNN approach. 

The manuscript has been organized as follows. In Sec.\,\ref{Formalism}
we review briefly the sensitivity of the structure of the outer crust of a
neutron star to nuclear masses and discuss in detail the Bayesian 
neural network approach to the calculation of masses. In
Sec.\,\ref{Results} we discuss the significant improvement to the mass
models after BNN refinement. Moreover, we used the newly developed
mass model to extract the composition of the stellar crust as a function
of depth. Finally, we conclude in Sec.\,\ref{Conclusions} with a summary
of the important findings and on future prospects to extend the BNN
formalism to other nuclear observables.


\section{Formalism}
\label{Formalism}

\subsection{The Physics of the Outer Crust}
\label{Outer Crust}

Although the most common perception of a neutron star is that of a uniform assembly 
of neutrons packed to enormous densities, the reality is far different and much more 
interesting. First, chemical equilibrium and charge neutrality favor a small but 
non-negligible fraction of protons and neutralizing electrons in the neutron star. Perhaps 
surprisingly, some of the fascinating phases that emerge in a neutron star are inextricably 
linked to the electrons. This is because the electronic Fermi energy increases rapidly with
density which drives the matter in the star to become neutron rich. Second, in hydrostatic
equilibrium, the pull by gravity on any mass element is exactly compensated by the gradient 
in the pressure. This implies, at least for ``conventional'' neutron stars, that the enormous 
pressure and density at the center of the star must both decrease to zero at the edge of the 
star. The enormous range of densities and extreme neutron-proton asymmetries are responsible 
for the many fascinating phases of a neutron star. 

In particular, at the very low densities of the outer crust a \emph{uniform} system of 
neutrons, protons and electrons is unstable against cluster formation. That is, at such low
densities the average inter-nucleon separation is significantly larger than the range of
the nucleon-nucleon interaction. Thus, it becomes energetically favorable for nucleons 
to cluster into nuclei that arrange themselves in a crystalline structure as a result of the 
long range Coulomb interaction. Although low for nuclear standards, at these densities
the neutralizing electrons have been pressure ionized and may be treated as a uniform
relativistic free Fermi gas\,\cite{Baym:1971pw}. The dynamics of the outer crust is 
thus encapsulated in the following simple expression for the total energy per 
nucleon\,\cite{Haensel:1989,Haensel:1993zw,Ruester:2005fm,RocaMaza:2008ja,RocaMaza:2011pk}:
 \begin{equation}
   \mathcal{E}(Z,A; n) = \frac{M(Z,A)}{A} + 
   \frac{m_{e}^{4}}{8\pi^{2}n} 
   \left[x_{{}_{\rm F}}y_{{}_{\rm F}}\Big(x_{{}_{\rm F}}^{2}+y_{{}_{\rm F}}^{2}\Big)
  -\ln(x_{{}_{\rm F}}+y_{{}_{\rm F}}) \right]
   -C_{l}\frac{Z^2}{A^{4/3}}\,p_{\rm F}.
\label{EperA}   
\end{equation}
The first term is independent of the baryon density of the system ($n\!=\!A/V$) and 
represents the entire nuclear contribution to the energy. It depends exclusively on 
the mass per nucleon of the nucleus populating the crystal lattice. The second term 
contains the contribution from a relativistic free Fermi gas of electrons of mass $m_{e}$, 
scaled Fermi momentum $x_{{}_{\rm F}}\!=\!p_{{}_{\rm F}}^{e}/m_{e}$, and scaled Fermi 
energy $y_{{}_{\rm F}}\!=\!\sqrt{1+x_{{}_{\rm F}}^{2}}$. The electronic Fermi momentum 
depends exclusively on the baryon density $n$ and the electron-to-baryon fraction $Z/A$:
 \begin{equation}
   p_{{}_{\rm F}}^{e} = (3\pi^2n_{e})^{1/3} = \left(3\pi^2n\frac{Z}{A}\right)^{1/3} =
   \left(\frac{Z}{A}\right)^{1/3}\!\!p_{{}_{\rm F}}.
\end{equation}
Finally, the last term provides the relatively modest---although by no means 
negligible---electrostatic lattice contribution ($C_{l}\!=\!3.40665\!\times\!10^{-3}$). It has 
a structure similar to the Coulomb term in the liquid drop formula [see Eq.\,(\ref{LDM})] 
but contributes with the opposite sign\,\cite{RocaMaza:2008ja}. In turn, the pressure of 
the system---which is dominated by the electronic contribution---is given at zero 
temperature by the following expression:
 \begin{equation}
   P(Z,A; n) = n^{2}\!\left(\frac{\partial\mathcal{E}}{\partial n}\right)_{T\!=\!0} =
   \frac{m_{e}^{4}}{3\pi^{2}} 
   \left(x_{{}_{\rm F}}^{3}y_{{}_{\rm F}} - 
   \frac{3}{8}\left[x_{{}_{\rm F}}y_{{}_{\rm F}}
   \Big(x_{{}_{\rm F}}^{2}+y_{{}_{\rm F}}^{2}\Big)
  -\ln(x_{{}_{\rm F}}+y_{{}_{\rm F}}) \right]\right)
  -\frac{n}{3}C_{l}\frac{Z^2}{A^{4/3}}\,p_{\rm F}.
 \label{Pressure}   
\end{equation}

Given that hydrostatic equilibrium demands that the ``optimal nucleus'' populating
the lattice be obtained at fixed pressure rather than at fixed density, the composition 
of the outer stellar crust is obtained by minimizing the chemical potential of the system. 
That is,
 \begin{equation}
   \mu(Z,A; P) = \frac{M(Z,A)}{A} + \frac{Z}{A}\mu_{e} -
   \frac{4}{3}C_{l}\frac{Z^2}{A^{4/3}}\,p_{\rm F} 
   \label{ChemPot}
\end{equation}
where $\mu_{e}$ is the electronic chemical potential. Note that the connection between 
the pressure and the baryon density is provided by the underlying crustal equation of 
state; see Eq.\,(\ref{Pressure}).

The search for the composition of the stellar crust is performed as follows. For a given 
pressure $P$ and nuclear species ($Z,A$), the equation of state is used to determine 
the corresponding baryon density of the system which, in turn, determines the Fermi 
momentum $p_{\rm F}$ and the electronic chemical potential $\mu_{e}$. Then, for each 
nuclear species one proceeds to compute the chemical potential $\mu(A,Z; P)$; this 
requires scanning over an entire mass table---which in some cases consists of nearly 
10,000 nuclei. Finally, the $(Z,A)$ combination that minimizes $\mu(A,Z; P)$ determines 
the nuclear composition of the crust at the given pressure. Naturally, if the density is very 
small so that the electronic contribution to the energy may be neglected, then 
${}^{56}$Fe---with the lowest mass per nucleon---becomes the nucleus of choice. (Note 
that whereas ${}^{56}$Fe has the lowest mass per nucleon it is ${}^{62}$Ni that has the 
largest binding energy per nucleon.)  As the pressure and density increase so that the 
electronic contribution may no longer be neglected, then it becomes advantageous to 
reduce the electron fraction $Z/A$. However, this can only be done at the expense of 
increasing the neutron-proton asymmetry which, in turn, results in an increase in the 
mass per nucleon. The question of which nucleus becomes the preferred choice then 
emerges from a competition between the electronic contribution (that favors $Z/A\!=\!0$) 
and the nuclear symmetry energy (which favors nearly symmetric nuclei). 

In summary, the structure of the outer stellar crust consists of a nuclear lattice embedded 
in an electron gas that is responsible for driving the system towards progressively more 
neutron rich nuclei. In this way, the outer crust represents a unique laboratory for the 
study of neutron-rich nuclei in the $Z\!\simeq\!20$-$50$ region that nicely complements 
our quest for a detailed map of the nuclear landscape at terrestrial laboratories. In the 
following section we introduce the BNN approach that will be used to predict the masses 
of the nuclei (some of them highly exotic) that populate the outer crust.

\subsection{Bayesian Neural Network Approach to Nuclear Masses}
\label{BNN}
Our basic idea is to view the modeling 
of $\delta_\textrm{LDM}(Z, N)$ in Eq.~(\ref{BNN_LDM}) as a problem of statistical inference of which there
are two main approaches: ``frequentist'' 
and ``Bayesian'', which differ in their interpretations of probability. 
Frequentists
consider probability to be a property of the physical world, whereas Bayesians
consider probability to be a measure of uncertainty regarding our knowledge 
of the physical world\,\cite{Stone:2013}. Consequently, in the  frequentist approach 
a probability can be assigned neither to
an hypothesis nor to a parameter whereas such assignments are natural
in the Bayesian context.  The cornerstone of our computational approach is a
Bayesian neural network (BNN),
a ``universal approximator" that is  capable,
in principle, of approximating any real function of one or more real 
variables\,\cite{Titterington:2004,Neal1996}.
The utility of the Bayesian
approach to neural networks  
is that it furnishes an estimate of the uncertainty in the approximated function in a computationally
convenient manner and it is less prone to overfitting that function\,\cite{Titterington:2004,Neal1996}.

The Bayesian approach to statistical inference is deeply rooted in Bayes' 
theorem, which provides a connection between a given set 
of data $D$ and a given hypothesis (or model) $H$\,\cite{Stone:2013},
\begin{equation}
 p(H|D)=\frac{p(D|H)p(H)}{p(D)}.
 \label{BayesRule}
\end{equation} 
The posterior probability $p(H|D)$ is the probability that the assumed
hypothesis is true given data $D$ and the prior probability  of the hypothesis $p(H)$.
 For example, given that a 
patient has tested positive for the ebola virus (empirical data $D$), what is the
probability that the patient has in fact contracted the disease (assumed
hypothesis $H$)? This question cannot be answered 
satisfactorily without specifying two
probabilities: the likelihood $p(D|H)$, which 
represents the probability that a patient that is actually known to be sick ($H$)
tests positive to ebola screening ($D$), and the prior probability of
being sick $p(H)$. Note that whereas $p(H|D)$ makes a 
statement about the well-being of the patient, $p(D|H)$ embodies the 
accuracy of the diagnostic test. The two are connected by Bayes' theorem 
as stated in Eq.\,(\ref{BayesRule}), with the connection provided by $p(H)$ 
(the probability of having ebola, say 1 in 10,000 among the population of 
Freetown in Sierra Leone during the 2014 epidemic) and $p(D)$ (the 
probability of testing positive).

The aim of the present work is to use Bayes' theorem to infer the 
probability that a given neural network model, defined by a set
of neural network model parameters,  describes a given 
set of experimental nuclear masses (empirical data). Using $(x,t) \equiv D$ to 
denote the relevant input and output data (see below) and $\omega \equiv H$ 
to denote the full set of model parameters, we write the posterior 
probability of interest as, 
\begin{equation}
 p(\omega|x,t)=\frac{p(x,t|\omega)p(\omega)}{p(x,t)},
 \label{BayesRule}
\end{equation} 
where  $p(x,t|\omega)$ is the likelihood and $p(\omega)$
is the prior density of the parameters $\omega$. Following standard practice, we assume a Gaussian 
distribution for the likelihood based on an objective (or ``loss'') function
obtained from a least-squares fit to the empirical data. That is,
\begin{equation}
 p(x,t|\omega)=\exp\big(\!-\chi^2/2\big),
 \label{likelihood}
\end{equation} 
where the objective function $\chi^{2}(\omega)$ is given by
\begin{equation}
 \chi^2(\omega)=\sum_{i=1}^{N}
 \left(\frac{t_i-f(x_i,\omega)}{\Delta t_{i}}\right)^2.
 \label{Chi2}
\end{equation}    
Here $N$ is the number of empirical data, $t_{i}\!\equiv\!t(x_{i})$ is the 
{\sl ith} observable with $\Delta t_{i}$ its associated error, and the 
function $f(x,\omega)$ (given below) depends on both the input data 
$x$ and the model parameters $\omega$. In our particular 
case, $x$ denotes the two input variables $x\!\equiv\!(Z,A)$ and 
$t(x)\!\equiv \delta_\textrm{LDM}(Z,A)$ is the mass residual. 

In the non-Bayesian approaches to neural networks, the function $\chi^2(\omega)$ is 
minimized to find a single best-fit value $\omega^*$ for the neural network parameters, and
hence a single best-fit neural network, $f(x, \omega^*)$.
However, rather than minimizing the objective function as it is 
conventionally done, we make predictions  by 
averaging the neural network over the posterior probability density of the network
parameters $\omega$, 
\begin{equation}
 \langle f_{n}\rangle = \int f(x_{n},\omega)p(\omega|x,t)\,d\omega
 =\frac{1}{K}\sum_{k=1}^{K} f(x_{n},\omega _{k}),
 \label{Avgfn}
\end{equation} 
where $x_{n} = (Z_n, A_n)$ are the parameters of the nucleus for which we wish
to predict the mass residual.  The high-dimensional integral in Eq.~(\ref{Avgfn}) is approximated by
Monte Carlo integration in which
the posterior probability $p(\omega|x, t)$ is sampled using the 
hybrid Markov Chain Monte Carlo (HMCMC) method~\cite{Neal1996}.
As noted above, an
enormous advantage of this approach is that it provides an estimate
\begin{equation}
 \Delta f_{n} = \sqrt{\langle f_{n}^{2}\rangle - \langle f_{n}\rangle^{2}},
 \label{Errorfn}
\end{equation} 
of the uncertainty in the theoretical prediction.

We now specify the form of the functions $f(x,\omega)$ and
$p(\omega)$. Note that, in principle, Bayes' theorem requires
specification of the function $p(x,t)$. However, since the MCMC method only
requires knowledge of the \emph{relative} posterior probabilities, the function 
$p(x,t)$ may be ignored.

\begin{figure}[ht]
\vspace{-0.05in}
\includegraphics[width=0.5\columnwidth, angle=0] {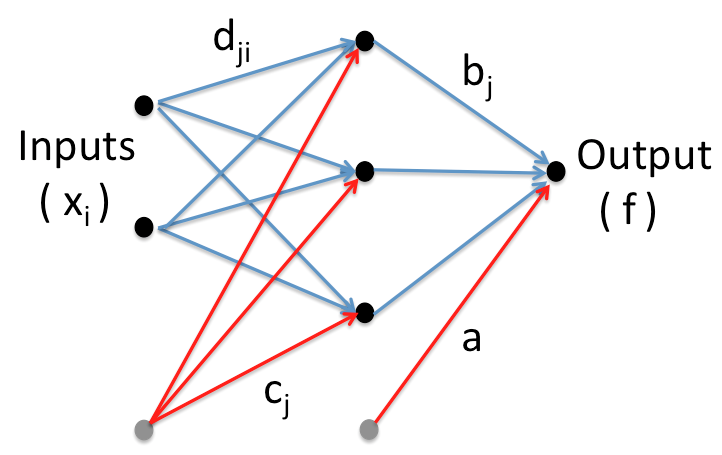}
\caption{A feed-forward neural network with a single hidden layer, 
 two inputs $Z$ and $A$, and a single output $f = \delta_\textrm{LDM}(Z,A)$.} 
\label{Fig1}
\end{figure}

In this work, we use a feed-forward neural network model defined by
\begin{equation}
  f(x,\omega)=a+\sum_{j=1}^H b_j {\rm tanh}\left(c_j+\sum_{i=1}^I d_{ji} x_i\right),
  \label{ANN}
\end{equation}
where the model parameters are given by $\omega\!=\!\big\{a,b_{j},c_{j},d_{ji}\big\}$, 
$H$ is the number of hidden nodes, and $I$ is the number of inputs. For 
two input variables ($Z$ and $A$), the function in Eq.~(\ref{ANN}) contains a total 
of $1\!+\!4H$ parameters. Since there are no \emph{a priori} criteria to decide the optimal 
number of hidden nodes $H$, some study is required to find the best choice. The
architecture of the neural network is shown in Fig.~\ref{Fig1}.

The specification of a prior is an essential part of any Bayesian analysis.  For this
problem, the prior density $p(\omega)$
should encode what is known about the neural network parameters. A priori, these parameters
can be positive or negative and, with the exception of parameter $a$ in Eq.~(\ref{ANN}), should be constrained to lie close to zero in order to obtain an approximation for $\delta_\textrm{LDM}(Z,A)$
that is as smooth as possible. We therefore follow Ref.~\cite{Neal1996} and assign a zero mean Gaussian
prior for each neural network parameter, while similar parameters in
Eq.~(\ref{ANN}) are assigned the same standard deviation: $\sigma_a$ for parameter $a$, $\sigma_b$ for the parameters
$b_j$, $\sigma_c$ for the parameters $c_j$, and $\sigma_d$ for the parameters $d_{ji}$. However,
since \emph{a priori}, we do not know what values should be assigned to these standard deviations,
we allow them to vary over a range by constraining the precisions ($1/\sigma^2$) using
a prior (which is often referred to as a hyperprior, that is, a ``prior that constrains a prior") for each of the four standard deviations, each modeled as a gamma density defined by two fixed parameters~\cite{Neal1996}.
The fixed parameters of the gamma densities are chosen so as to maximize 
the accuracy of the predictions. The prior $p(\omega)$ is therefore the integral with respect to the
four precisions of
a product of Gaussians, one
for each neural network parameter, times the four gamma densities, one for each of the 
precisions, $1/\sigma_a^2$, $1/\sigma_b^2$, $1/\sigma_c^2$, and $1/\sigma_d^2$.

Having laid out the foundation of the BNN method, we now proceed to construct a model of nuclear 
masses by training BNNs on mass 
residuals, 
\begin{equation}
 t(x) = M_{\rm exp}(x) - M_{\rm th}(x),
 \label{residuals}
\end{equation}
that is, the difference between the experimental values and the theoretical predictions 
from a given mass model. 
This strategy is consistent with the approach articulated in the Introduction: we include as 
much physics as possible by using the physics-motivated models in the
literature and use the BNN to fine tune these models by modeling the residuals.


\section{Results}
\label{Results}

To illustrate the BNN approach we begin with the simplest mass model available in the literature:
the liquid drop model of Bethe and Weizs\"acker introduced in Eq.\,(\ref{LDM}). As it is customarily
done, optimal values for the six empirical parameters are determined from a least-squares fit to 
the experimental binding energies of the more than 3,200 nuclei listed in the latest AME2012 
compilation\,\cite{AME:2012}. Note that by implementing a MCMC-Metropolis algorithm for a
likelihood function defined as in Eq.(\ref{likelihood})\,\cite{Piekarewicz:2014kza}, one can 
obtain optimal values with associated theoretical uncertainties; see Table\,\ref{Table1}.

\begin{center}
\begin{table}[h]
\begin{tabular}{|c|c|c|c|c|c|}
\hline  $a_{\rm v}$(MeV)  & $a_{\rm s}$(MeV) & $a_{\rm c}$(MeV) & $a_{\rm a}$(MeV) 
      & $a_{\rm as}$(MeV) & $a_{\rm p}$(MeV) \\
\hline $15.422(17)$& $16.831(53)$& $0.686(1)$& $26.002(111)$& $-18.711(482)$& $11.199(388)$\\
\hline
\end{tabular}
\caption{Liquid-drop-model parameters and uncertainties obtained from the latest AME2012 
 compilation of nuclear masses\,\cite{AME:2012}.} 
\label{Table1}
\end{table}
\end{center}

Having defined a theoretical model one can now start with the implementation of the BNN 
algorithm. The training of the neural network requires a separation of the data into three 
different sets:  (a) learning, (b) validation, and (c) prediction. The learning set consists of 
a randomly selected group of nuclei within the AME2012 database that will be used to sample
the parameters of the neural network function given in Eq.(\ref{ANN}). The validation 
set comprises nuclei that are still within the AME2012 database but that were not used 
in the modeling of the residual function $\delta_\textrm{LDM}(Z,A)$. 
Finally as the name suggests, the prediction 
set consists of a group of nuclei not contained in the AME2012 compilation but that are 
vital for elucidating phenomena sensitive to such (unknown) masses, as in the case of 
the composition of the neutron star crust. 

In the spirit of Strutinsky's energy theorem\,\cite{Strutinsky:1967}, we assume that the
liquid drop model provides---as indeed it does---an accurate description of the large and 
smooth behavior of the underlying mass function. Then, the BNN algorithm is used to
refine the LDM predictions by modeling $\delta_\textrm{LDM}(Z,A)$. 
In the case of the LDM, the 
residuals represent the small deviations that are not captured by the LDM model.
To avoid regions of the nuclear landscape where the masses fluctuate too rapidly (as in
the case of light nuclei) or where the experimental uncertainties are large (such as 
for very massive nuclei), we limit our data set to the 2591 nuclei between $^{40}$Ca 
and $^{240}$U. From this limited (yet still very large) set, the learning set is built
from 1800 randomly selected nuclei (about 70\% of the original set). The remaining 
791 nuclei constitute the validation set. With two input variables ($Z$ and $A$) and 
$H\!=\!40$ hidden nodes, a total of $1\!+\!4H\!=\!161$ parameters must be sampled. 
To do so, we use the Flexible Bayesian Modeling package by Neal described in 
Ref.~\cite{Neal1996}. After an initial thermalization phase consisting of 500 iterations, 
sampling data are accumulated for a total of 100 iterations that are used to determine statistical 
averages, via Eq.~(\ref{Avgfn}), and their associated uncertainties. 

To assess the quality of the resulting neural network function $f(x,\omega)$, we 
compute the mean-square deviation 
\begin{equation}
 \sigma^{2} = \frac{1}{K}\sum_{k=1}^{K} 
 \Big[M_{\rm exp}(k)-M_{\rm th}(k)\Big]^{2},
 \label{MSdeviation}
\end{equation}
of the mass of the $K\!=\!290$ nuclei (out of the 791 nuclei in the validation set) 
that are of relevance to the composition of the outer stellar crust, namely, those 
spanning the $Z\!=\!20$-$50$ region. Note that in the above expression
``exp'' stands for the experimentally quoted value in the AME2012 compilation and 
``th'' for the corresponding theoretical prediction. The root-mean-square deviation as 
per Eq.\,(\ref{MSdeviation}) for a representative set of sophisticated mass models are 
displayed in Table\,\ref{Table2}. These include the microscopic-macroscopic mass 
models of Duflo and Zuker (DZ)\,\cite{Duflo:1995}, M\"oller and 
Nix (MN)\,\cite{Moller:1981zz,Moller:1988}, and the finite range droplet 
model (FRDM)\,\cite{Moller:1993ed}, alongside the two accurately calibrated 
microscopic models HFB19 and HFB21\,\cite{Goriely:2010bm}. 

As shown in Table\,\ref{Table2}, for all these five mass models the root mean 
square deviation---denoted as $\sigma_{\rm pre}$---falls in the range of 
$0.5$-$1$\,MeV. In contrast and consistent with expectations, the simple liquid 
drop model yields a deviation that is considerably larger ($\sim\!3.6\,{\rm MeV}$). 
However, once properly trained, the BNN-improved liquid-drop model 
(listed on the second line as $\sigma_{\rm post}$) rivals the predictions of the 
most accurate of these models. This important finding validates the basic tenet
of this work, namely, that the small and fluctuating contribution to the nuclear 
mass may be accounted for by properly training on the residuals.

\begin{center}
\begin{table}[h]
\begin{tabular}{|l||c|c|c|c|c|c|}
 \hline
 Model & LDM & DZ  & MN & FRDM & HFB19 & HFB21 \\
 \hline
 $\sigma_{\rm pre}$\,(MeV)   & 3.359 & 0.526 & 0.963 & 0.861 & 0.880 &  0.816   \\
$\sigma_{\rm post}$\,(MeV)   & 0.556 & 0.303 & 0.507 & 0.460 & 0.524 &  0.555  \\
\hline
$\Delta\sigma/\sigma_{\rm pre}$    & 0.835 & 0.424 & 0.474 & 0.466 & 0.405 & 0.320  \\

\hline
\end{tabular}
\caption{Root-mean-square deviation as predicted by a representative set of
models for the mass of 290 nuclei of possible relevance to the outer crust of 
a neutron star; see text for details.}
\label{Table2}
\end{table}
\end{center}

\begin{figure}[h]
\vspace{-0.05in}
\includegraphics[width=0.5\columnwidth,angle=0]{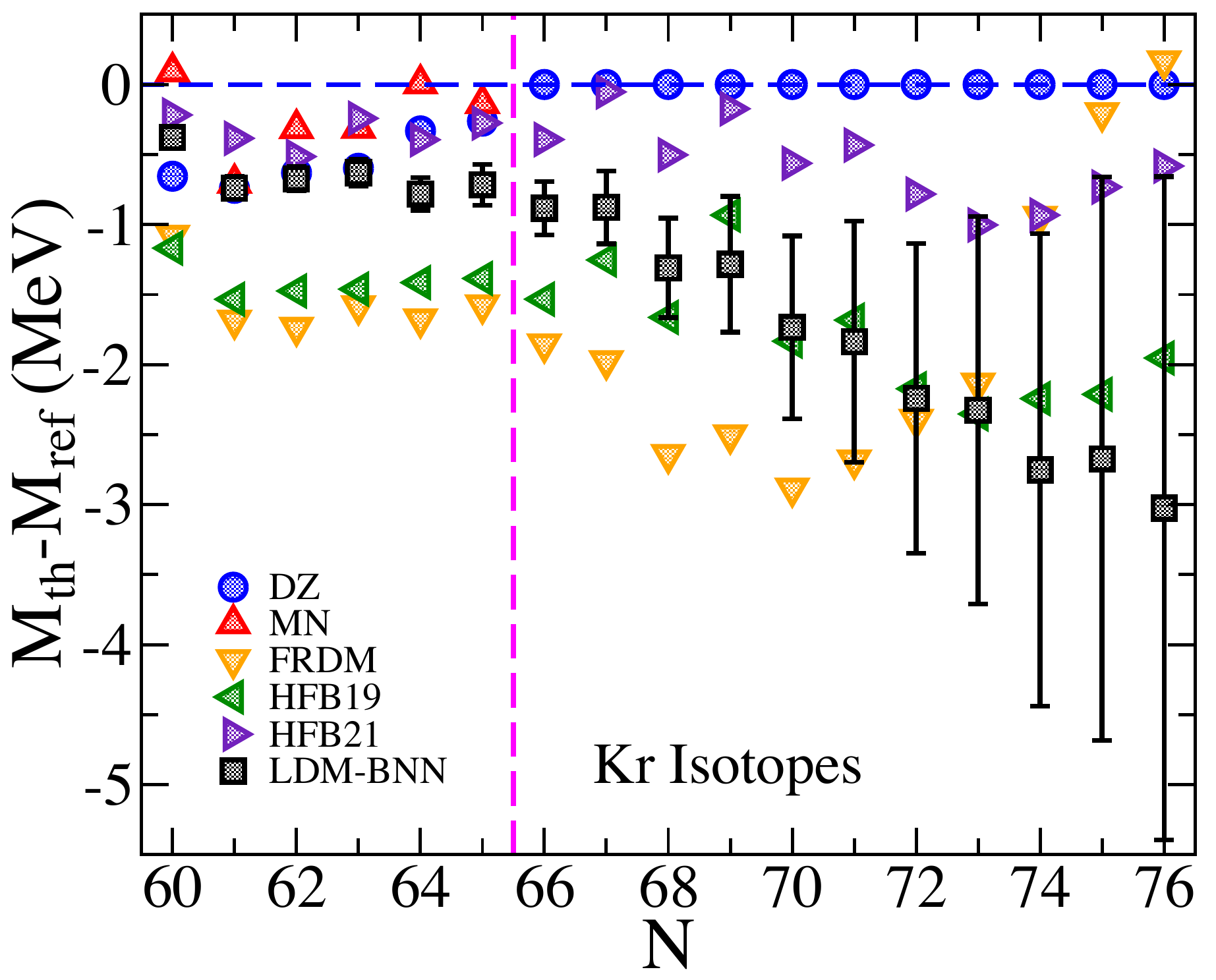}
\caption{Mass predictions for the Krypton isotopes relative to a 
reference mass from all the five mass models considered in the 
text. Also shown are the predictions from the BNN-improved liquid 
drop model with its associated theoretical errors. The reference
mass is taken from AME2012 for $60\!\le\!N\!\le\!65$ and from
the Duflo-Zuker model for $N \!>\!65$.}
\label{Fig2}
\end{figure}

For a graphical depiction of our findings--and with an eye on further 
refinements---we display in Fig.\,\ref{Fig2} predictions for the masses
of the Krypton isotopes ($Z\!=\!36$) in the ${}^{96-112}$Kr range. For
ease of viewing, we plot the theoretical predictions relative to a 
reference mass. For the ${}^{96-101}$Kr region where experimental
masses are available, we use the AME2012 tabulated values\,\cite{AME:2012},
whereas for the $N\!\ge\!66$ region we use the Duflo-Zuker predictions 
as the reference mass; this ``transition'' region is delineated by the dashed 
vertical line. Besides predictions from the five models (DZ, MN, FRDM,
HFB19, and HFB21) we include BNN-improved results from the liquid drop 
model with associated theoretical uncertainties. Having previously validated 
the BNN algorithm, these predictions were made with a refined neural network 
function that used as the learning set all 2591 nuclei between $^{40}$Ca 
and $^{240}$U.

In the $60\!\le\!N\!\le\!65$ region, the predictions of all the models---including 
the BNN improved LDM---are within 2 MeV of the experiment. Perhaps more 
relevant is the fact that the statistical errors associated with the 
LDM-BNN predictions suggest that in this region the \emph{systematic} errors 
associated with the various models (although relatively small) dominate over 
the statistical uncertainties. This indicates a need for a better understanding of 
the sources that dominate the $\sim\!2$\,MeV systematic uncertainties. 
In sharp contrast, the uncertainties in the $N \!>\!65$ region where no data
is available are dominated by the statistical error---especially for the most 
neutron-rich isotopes. Without errors, one could be under the false impression 
that the models are inconsistent with each other. This fact underscores the 
critical importance of uncertainty quantification. Indeed, theoretical predictions 
without accompanying statistical errors---especially when large extrapolations 
are involved---are of very little value. Finally, our results highlight the vital role
of future rare isotope facilities. Although the outer crust requires extrapolations 
into regions of the nuclear chart that are unlikely to be explored even with the 
most sophisticated rare isotope beam facilities---after all, ${}^{118}$Kr is 21 
neutrons away from the last isotope with a well measured mass---mass 
measurements of even a few of these exotic short-lived isotopes could prove
crucial in informing nuclear-structure models. 

\begin{figure}[h]
\vspace{-0.05in}
\includegraphics[width=0.5\columnwidth,angle=0]{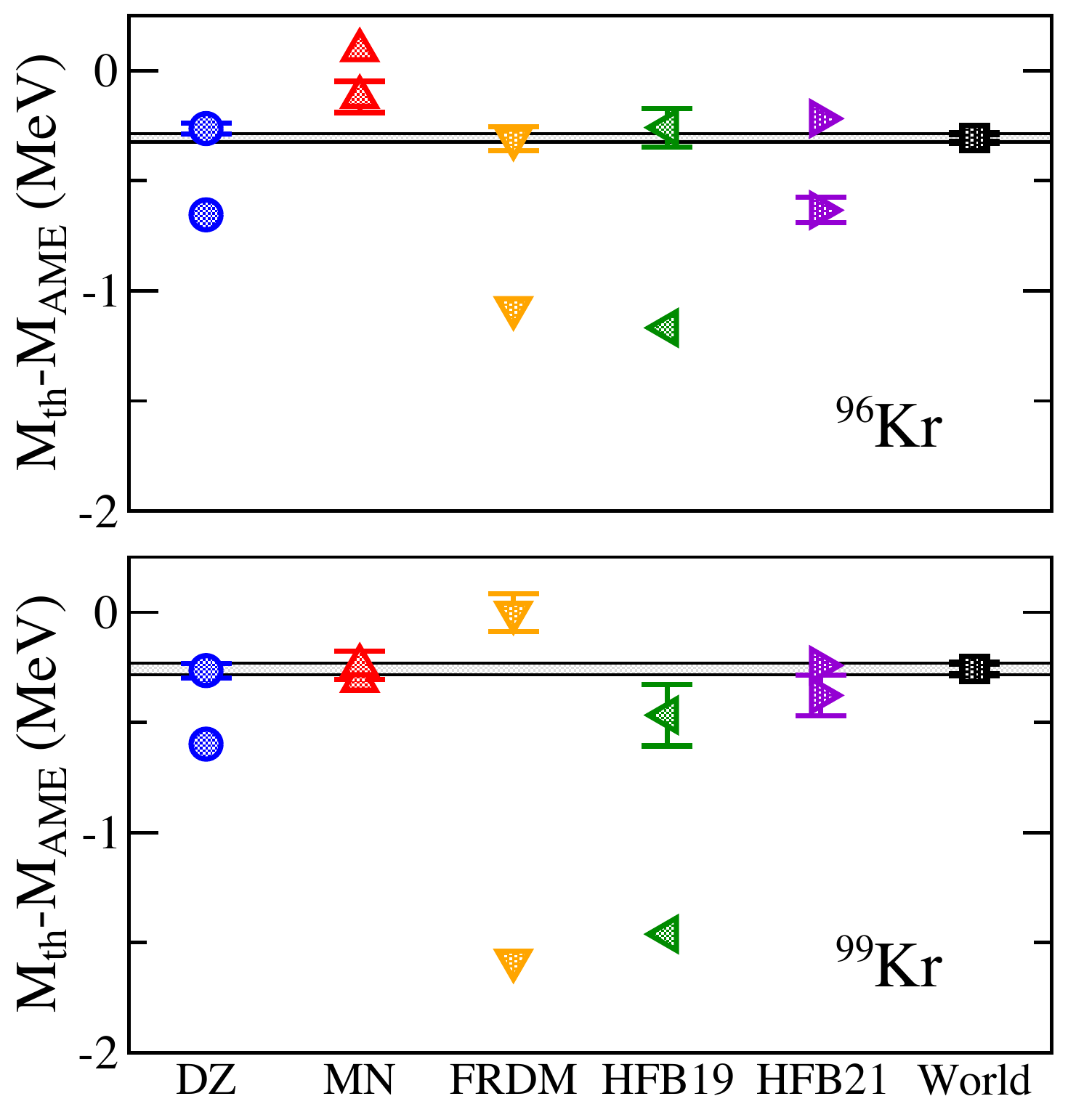}
\caption{Pre- and post-BNN improved mass predictions relative to 
the AME2012 tabulated values for ${}^{96}$Kr and ${}^{99}$Kr. The
BNN predictions include statistical errors and ``World'' represents the 
world average of the five models obtained as per Eq.\,(\ref{WorldAvg}).}
\label{Fig3}
\end{figure}

Given the promise of the approach, it seems natural to extend the BNN formalism 
to the five high-quality mass models considered in this work. Thus, exactly as it was
done in the case of the liquid-drop model, we use approximately 70\% of the nuclear 
masses tabulated in the AME2012 compilation to train (using mass residuals) each 
of the five individual mass models. What emerges are five different neural network 
functions each with its own set of parameters. Once calibrated, we then use the
same 290 nuclei (out of 791) that were used earlier to validate the LDM-BNN model 
to assess the quality of the BNN refinement. The resulting root-mean-square deviation 
$\sigma_{\rm post}$ are listed in Table\,\ref{Table2} alongside the previously shown 
result for the liquid drop model. In all cases we observe a considerable improvement.
This is particularly significant given that these represent some of the most sophisticated
mass models available to date. This observation validates our approach of 
incorporating as much 
physics as possible into the underlying mass model but ultimately relying on an
empirical BNN model to refine the mass model.

\begin{figure}[h]
\vspace{-0.05in}
\includegraphics[width=0.4\columnwidth,height=8.5cm]{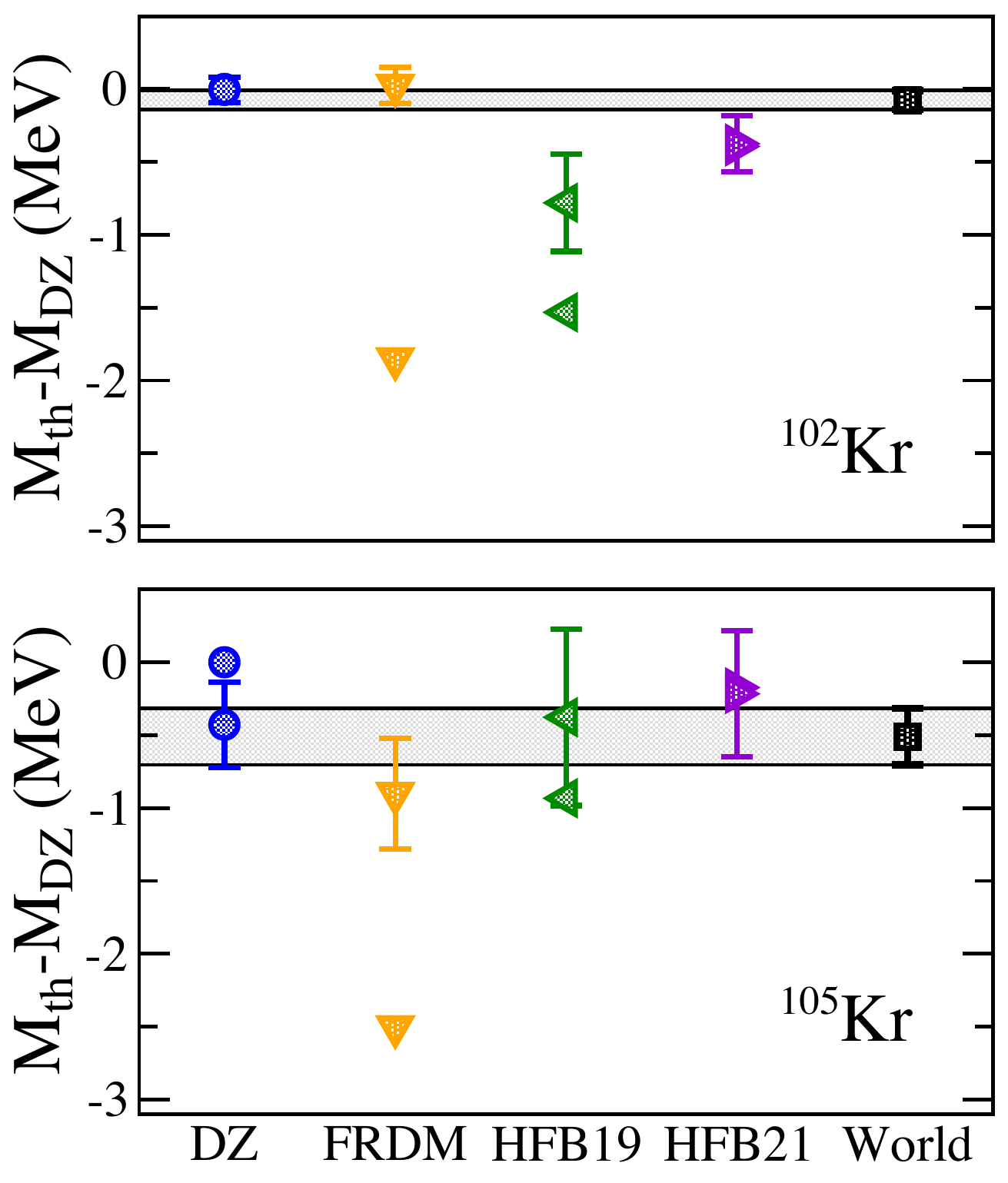}
\hspace{5pt}
\includegraphics[width=0.4\columnwidth,height=8.56cm]{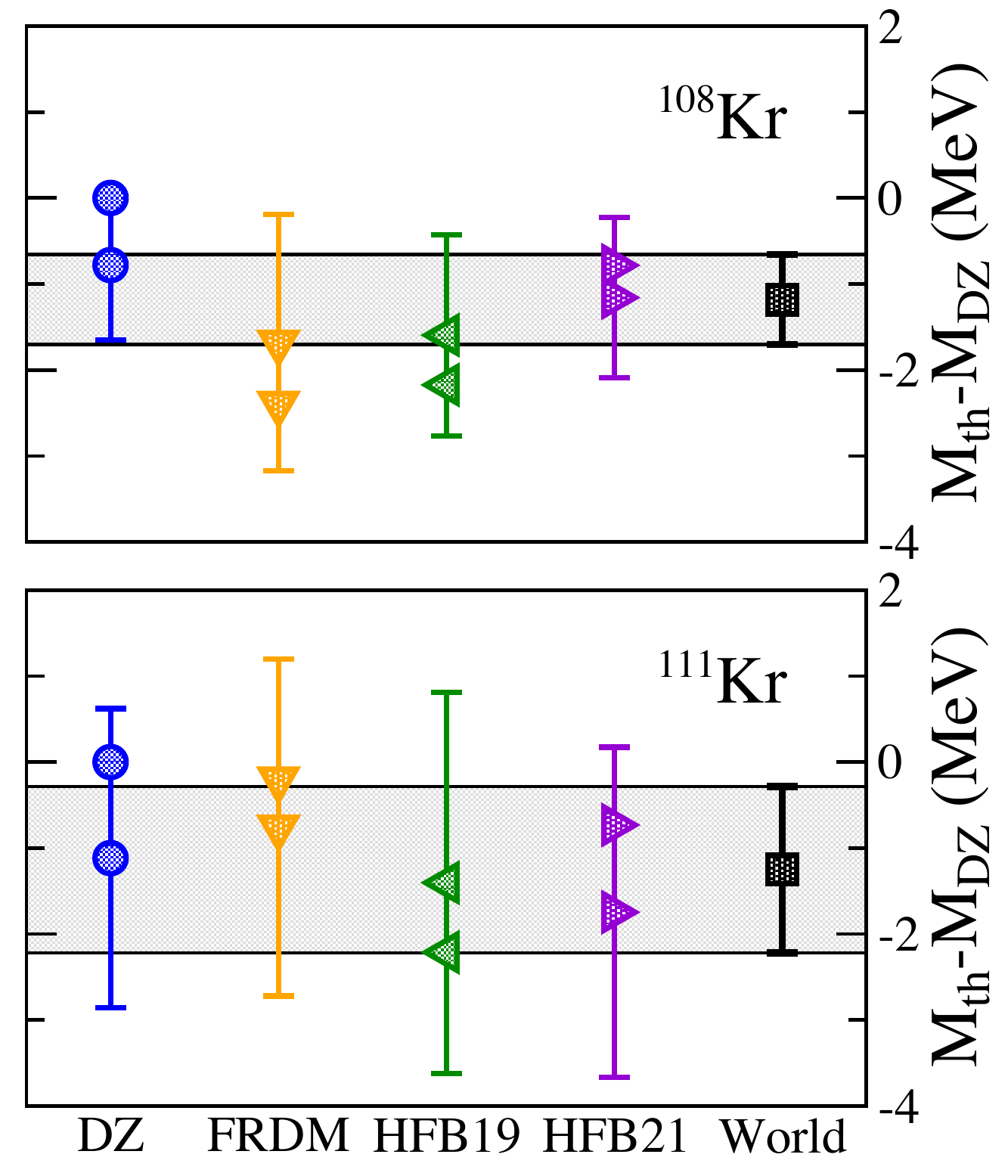}
\caption{Pre- and post-BNN improved mass predictions relative to 
the ``bare'' Duflo-Zuker values for ${}^{102}$Kr, ${}^{105}$Kr, ${}^{108}$Kr, 
and ${}^{111}$Kr. The BNN predictions include statistical errors and ``World'' 
represents the world average of the four models obtained as per 
Eq.\,(\ref{WorldAvg}).}
\label{Fig4}
\end{figure}

To illustrate this refinement in graphical form we display in Fig.\,\ref{Fig3} theoretical 
predictions for the masses of ${}^{96}$Kr and ${}^{99}$Kr relative to the experimental 
value\,\cite{AME:2012}. As in the case of Fig.\,\ref{Fig2}---and because extrapolations
are unavoidable---these predictions have been done using the entire AME2012 mass
compilation as the learning set. Although the pre-BNN predictions of all five models are 
fairly accurate, they display a significant amount of systematic variations. However, 
once the BNN refinement is implemented, most of these systematic differences 
disappear. Moreover, an estimate of uncertainty is now associated with each 
mass model. 
Ultimately, this enables us to compute a ``world average'' value by combining the 
BNN-improved predictions in the following way:
\begin{equation}
  M_{\rm world}=\sum_{n} \omega_{n}\,M_{n}\,, 
   \hspace{8pt}
  V_{\rm world}=\sum_{n} \omega_{n}^{2}\,V_{n}\,, 
   \hspace{4pt} {\rm and} \hspace{6pt}  
  \omega_{n} = \frac{V_{n}^{-1}}{\sum_{n}V_{n}^{-1}}\,,
 \label{WorldAvg} 
\end{equation}
where the sum runs over all the models and $V_{n}$ represents the variance 
of each model. As was done in Fig.\,\ref{Fig3}, we display in Fig.\,\ref{Fig4} the same 
trends but now for the case of the more exotic ${}^{102}$Kr, ${}^{105}$Kr, ${}^{108}$Kr, 
and ${}^{111}$Kr isotopes where experimental information is not yet available (also 
unavailable are predictions from the model by M\"oller-Nix). Given the lack of 
experimental data, the increase with $N$ of both the systematic and statistical uncertainties 
is hardly surprising. Again, this underscores the pressing 
need for measuring masses of exotic 
nuclei at rare isotope facilities. 

Having obtained a mass model---generated from the world averages as defined 
in Eq.\,(\ref{WorldAvg})---we are now in a position to predict the composition of 
the outer stellar crust. To do so, the pressure $P(r)$ and mass $M(r)$ profiles 
of the star are generated via the Tolman-Oppenheimer-Volkoff (TOV) equations:
\begin{align}
& \frac{dP}{dr} = -G \frac{M(r) \mathcal{E}(r)}{r^2} 
 \left[1 + \frac{P(r)}{\mathcal{E}(r)} \right] 
 \left[1 + \frac{4 \pi r^3 P(r)}{M(r)} \right]
 \left[1 - \frac{2 G M(r)}{r} \right]^{-1}\hspace{-12pt}, \\
 & \frac{dM}{dr} = 4 \pi r^2 \mathcal{E}(r).
\end{align}
Here $\mathcal{E}(r)$ is the energy density that is connected to the pressure
$P(r)$ via an equation of state. To illustrate the procedure we consider a 
``canonical'' $M_{0}\!=\!1.4\,M_{\odot}$ neutron star with a radius of 
$R_{0}\!=\!12.78$\,km as predicted by a realistic equation of 
state\,\cite{Todd-Rutel:2005fa}. These two quantities are sufficient to define 
the boundary conditions at the edge of the outer crust, namely, 
$M(R_{0})\!=\!M_{0}$ and $P(R_{0})\!=\!P_{0}\!\approx\!0$. Given $P_{0}$, 
the corresponding baryon density, energy density, and composition may be 
determined from the minimization of the chemical potential; see 
Eqs.\,(\ref{EperA}), (\ref{Pressure}), and (\ref{ChemPot}). At such 
an infinitesimal pressure (and baryon density), the crystalline lattice is 
composed of ${}^{56}$Fe nuclei. 

Knowledge of $M_{0}$, $P_{0}$ and $\mathcal{E}_{0}\!=\!\mathcal{E}(R_{0})$ 
is all that is needed to integrate inward the TOV equations to determine both 
the pressure and enclosed mass at the next (interior) point. With such pressure 
at hand, one proceeds once more to determine the associated baryon density, 
energy density, and composition at the given depth. This allows one to integrate 
inward the TOV equations to the next point, and so on. This iterative procedure 
continues until the total chemical potential of the system becomes equal to the 
free neutron mass. At this density it is no longer possible to bind all the neutrons 
into nuclei; the ``neutron drip line'' is reached. This stellar depth demarcates the 
transition from the outer to the inner crust.

\begin{figure}[ht]
\vspace{-0.05in}
\includegraphics[width=.8\columnwidth,angle=0]{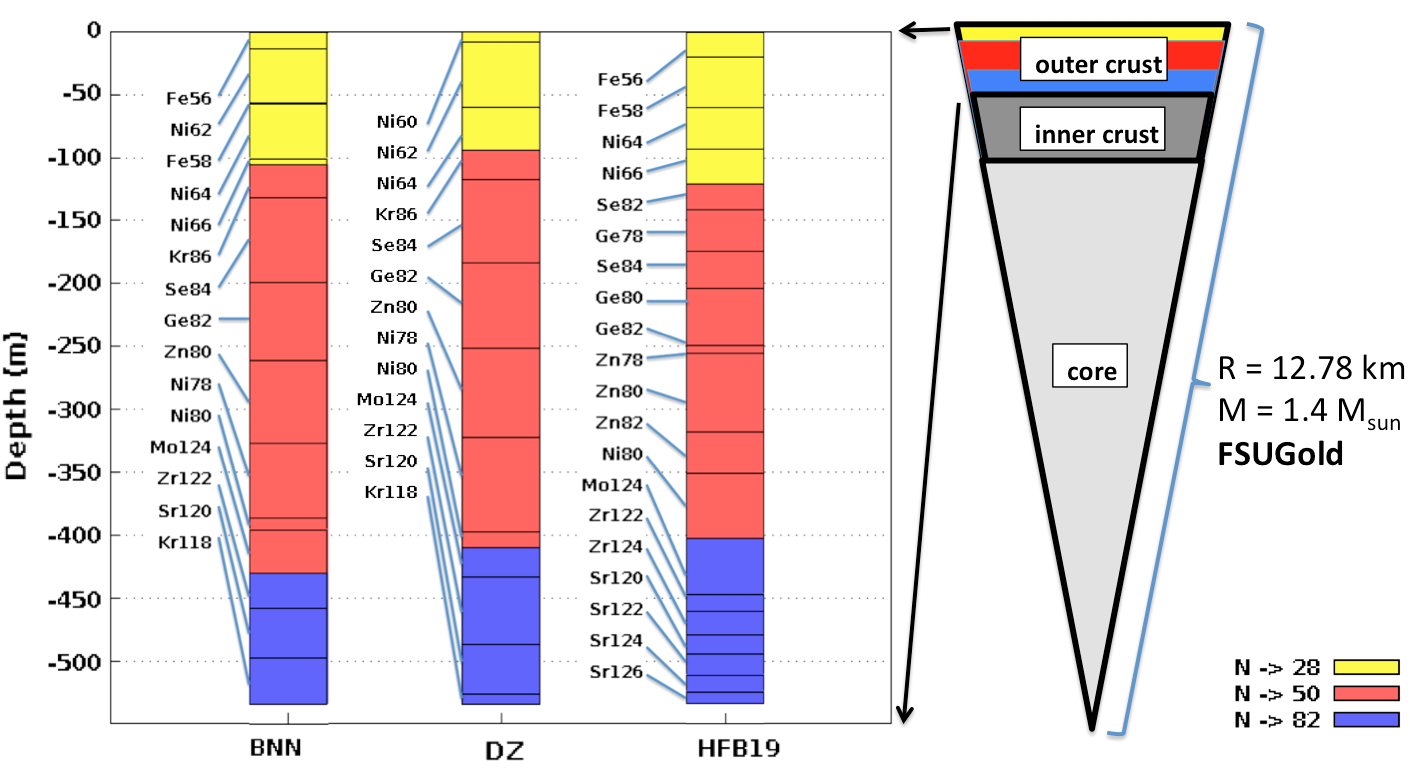}
\caption{Composition of a canonical $1.4\,M_{\odot}$ neutron
star with a 12.78\,km radius as predicted by three mass models:
``BNN-world", DZ, and HFB19.} 
\label{Fig5}
\end{figure}

In Fig.\,\ref{Fig5} we display the composition of the outer crust as a
function of depth for a neutron star with a mass of $1.4\,M_{\odot}$ 
and a radius of 12.78\,km. Predictions are shown using our newly 
created mass model ``BNN-world'', Duflo Zuker, and HFB19; these 
last two without any BNN refinement. The composition of the 
upper layers of the crust (spanning about 100 m and depicted in 
yellow) consists of Fe-Ni nuclei with masses that are well known 
experimentally. As the Ni-isotopes become progressively more 
neutron rich, it becomes energetically favorable to transition into 
the magic $N\!=\!50$ isotone region. In the particular case of 
BNN-world, this intermediate region is 
predicted to start with stable ${}^{86}$Kr and then progressively 
evolve into the more exotic isotones ${}^{84}$Se ($Z\!=\!34$), 
${}^{82}$Ge ($Z\!=\!32$), ${}^{80}$Zn ($Z\!=\!30$), and 
${}^{78}$Ni ($Z\!=\!28$); all this in an effort to reduce the electron 
fraction. In this region, most of the masses are experimentally known, 
although for some of them the quoted value is not derived from purely 
experimental data\,\cite{AME:2012}. Ultimately, it becomes energetically 
favorable for the system to transition into the magic $N\!=\!82$ isotone 
region. In this region \emph{none of the relevant nuclei have 
experimentally determined masses}. Although not shown, it is interesting
to note that the composition of the HFB19 model changes considerably
after the BNN refinement, bringing it into closer agreement with the
predictions of both BNN-world and Duflo-Zuker. Although beyond the 
scope of this work, we should mention that the crustal composition 
is vital in the study of certain elastic properties of the crust, such as its 
shear modulus and breaking strain---quantities that are of great 
relevance to magnetar starquakes\,\cite{Piro:2005jf,Steiner:2009yg} 
and gravitational wave emission\,\cite{Horowitz:2009ya}.

\section{Conclusions}
\label{Conclusions}

The determination of nuclear masses lies at the core of Nuclear Physics.
Starting almost eight decades ago with the pioneering work of Bethe and
Weizs\"acker and continuing to this day with the development of ever
more sophisticated theoretical models, the prediction of nuclear masses
is not only of great intrinsic interest but, in addition, plays a fundamental
role in elucidating a variety of astrophysical phenomena. However, despite 
the sophistication and success of modern mass models, systematic 
uncertainties associated with the constraints and limitations of each model 
remain. Moreover, these systematic uncertainties continue to grow as the 
models are extrapolated to uncharted regions of the nuclear landscape. 
Given that mass-sensitive astrophysical phenomena, such as r-process 
nucleosynthesis and the composition of the neutron star crust, demand 
knowledge of nuclear masses far away from stability, it becomes imperative 
to reconcile some of these differences. In this work we have introduced a
novel approach firmly rooted in Strutinsky's energy theorem that suggests 
that the nuclear binding energy may be separated into a large and smooth 
component and another one that is small and fluctuating. Using the liquid
drop model as an example to generate the large and smooth component,
we then invoked a Bayesian neural network approach to account for the
small and fluctuating component of the binding energy. The BNN formalism
is an approximation method that relies on the application 
of Bayes' theorem 
and a highly non-linear neural network function. 
By doing so, we obtained a refined 
LDM that rivals the predictions of the most sophisticated mass models 
available to date.

Motivated by the success of the BNN approach, we have extended the 
formalism to five of the most successful mass models available in the 
literature. The aim was to overcome the unavoidable limitations of any
model by building an artificial neural network function that could account 
for the small deviations from experiment. Moreover, due to the probabilistic 
nature of the Bayesian approach, the improved predictions were now
accompanied by proper theoretical errors. Despite the undeniable quality 
of the original mass models, significant improvements were observed in
all cases after the implementation of the BNN protocol. As important, the
spread among the various models was considerable reduced. Ultimately, 
a new mass model was obtained by combining the various model 
predictions (after BNN refinement) into a ``world average''. 

As a first test of the new mass model we have computed the composition 
of the outer crust of a neutron star, as it is only sensitive to nuclear masses 
in the $20\!\lesssim\!Z\!\lesssim\!50$ range. Whereas the composition in the
upper layers of the crust is model independent, the situation is drastically 
different in the high density layers where the models predict a composition 
that is unlikely to ever be recreated in the laboratory. Indeed, the exotic
nucleus of ${}^{118}$Kr---21 neutrons removed from the last isotope with 
a well measured mass---is predicted to lie at the very bottom layer of the
outer crust. Although mass measurements of some of these exotic 
$N\!=\!82$ isotones (such as ${}^{118}$Kr, ${}^{120}$Sr, ${}^{122}$Zr, 
and ${}^{124}$Mo) may not be feasible even at future state-of-the-art 
facilities, it is critical to continue this quest as far as possible from 
stability to properly inform theoretical models. 

The study of the composition of the stellar crust represents a proof-of-principle 
implementation of the BNN protocol to the important case of nuclear masses. 
However, this relatively simple example represents the ``tip of the iceberg''. 
For example, the newly created mass model may also be used to compute 
neutron separation energies for the neutron-rich isotopes of relevance to 
r-process nucleosynthesis. Moreover, the BNN framework is flexible and 
powerful enough to be extended to other physical observables. The basic 
requirement is the existence of a robust theoretical model with a strong 
physics underpinning, so that the residuals between theory and experiment 
become a smooth function of the input parameters ({\sl e.g.,} $Z$ and $A$). 
In that case, such a smooth function could be accurately represented by an 
artificial neural network function. Natural extensions of the BNN approach to 
other nuclear observables with already large experimental databases are 
charge radii and beta-decay lifetimes, among others. Work along these lines 
is currently in progress.


\begin{acknowledgments}
 We are very grateful to Dr. Michelle Perry for many fruitful discussions and for her 
 guidance into the subtleties of the Bayesian Neural Network approach. This material 
 is based upon work supported by the U.S. Department of Energy Office of Science, 
 Office of Nuclear Physics under Award Number DE-FD05-92ER40750.
\end{acknowledgments}


\bibliography{./BNNonNuclearMass.bbl}

\end{document}